\documentclass[sigconf]{acmart}

\AtBeginDocument{%
  }

\usepackage{dirtytalk}
\usepackage{csquotes}

\renewcommand\footnotetextcopyrightpermission[1]{} 

\setcopyright{none}

\acmConference[SBES 2026]{40th Brazilian Symposium on Software Engineering}{September 8–12, 2026}{São Paulo, SP, Brazil}

\begin{document}

\title{A Preliminary Study on the Impact of AI in the Creativity and Collaboration in Software Teams}
\titlenote{Pre-print of a publication from the 40th Brazilian Symposium on Software Engineering (SBES 2026)}

\author{Danilo Monteiro Ribeiro}
\email{ dmr@cin.ufpe.br}
\orcid{0000-0001-7393-729X}
\affiliation{%
  \institution{Universidade Federal de Pernambuco (UFPE), cesar.school}
  \city{Recife}
  \country{Brazil}
}
\author{Kiev Gama}

\email{kiev@cin.ufpe.br}
\orcid{0000-0003-1508-6196}
\affiliation{%
  \institution{Universidade Federal de Pernambuco (UFPE)}
  \city{Recife}
  \country{Brazil}
}

\author{Victoria Jackson}
\orcid{0000-0002-6326-931X}
\email{V.jackson@soton.ac.uk}

\affiliation{%
  \institution{University of Southampton}
  \city{Southampton}
  \country{UK}
  }


\begin{abstract}
Generative artificial intelligence (GenAI) tools are reshaping software engineering work, increasingly acting as cognitive partners rather than purely instrumental aids. Most AI tooling remains optimized for individual use, yet software development is inherently collaborative and creative, raising open questions about how AI affects team-based work. This paper examines how software engineering professionals perceive and integrate AI tools focusing on creativity and collaboration within teams. We conducted semi-structured interviews with 13 software professionals from four companies, each representing a distinct team. Our analysis identifies three central themes — AI Use, Consequences of AI, and Collaboration and Team Dynamics — alongside Emotions as a cross-cutting dimension. Key findings include: AI broadens developers' creative repertoires but may narrow independent ideation; developers increasingly consult AI instead of colleagues, weakening peer learning and mentoring; and teams are developing emerging triadic collaboration patterns in which developers, colleagues, and AI reason together. 
\end{abstract}



\maketitle

\section{Introduction}

In recent years, advances in Generative Artificial Intelligence (GenAI) tools have profoundly transformed Software Engineering (SE) practices. Language models and code assistants such as GitHub Copilot, ChatGPT, and Cursor are no longer merely support instruments; they increasingly act as collaborative partners capable of generating and refactoring code, suggesting solutions, and explaining errors~\cite{meem_ChatGPT_2025, seeber2020machines}. This shift has delineated a new frontier of investigation into human–AI collaboration, in which systems move beyond a purely instrumental role and become agents integrated into individual and team workflows. Recent studies suggest that transforming AI from an individual tool into a team member requires rethinking roles, orchestration, and work processes~\cite{seeber2020machines}, and that the nature of interactions among developers is already changing as a result~\cite{salomon_GenAIHumanInteractions_2026}. Throughout this paper, we use \textit{AI} and \textit{GenAI} interchangeably to refer specifically to generative AI tools (e.g., ChatGPT, GitHub Copilot); other forms of AI were not part of this study's scope.

Software development encompasses a wide range of creative activities, including feature ideation, interface design, architectural design~\cite{groeneveld_creativity_2021}, and coding~\cite{inman_developer_2024}. Creativity can be understood as the ability to produce ideas that are both new and valuable~\cite{boden2004creative}, and it is central to innovation in software teams, their processes, and products~\cite{gama_Startups_Creativity_2025}. AI tools have the potential to expand developers' creative repertoire~\cite{jackson_genai_creativity_24}, but they also introduce cognitive and ethical challenges, such as overreliance on algorithmic suggestions and dilution of authorship~\cite{amershi2019guidelines}. Creativity and collaboration are often intertwined, with creative work occurring both in collaborative settings such as whiteboarding~\cite{groeneveld_creativity_2021} and in independent work~\cite{inman_developer_2024}.

Although prior work has examined GenAI's effects on individual developer productivity~\cite{jetbrains_Survey_2025, meem_ChatGPT_2025, pereira_exploring_2025} and on changes in developer interactions~\cite{salomon_GenAIHumanInteractions_2026}, creativity and collaboration have been studied largely in isolation. This separation is a gap: in software teams, creative activities such as ideation, architectural reasoning, and design are inherently collaborative~\cite{groeneveld_creativity_2021}, and AI tools that reshape how developers interact necessarily reshape how they create together. Studying these two dimensions jointly allows us to surface tensions and trade-offs that neither lens alone can reveal, for instance, whether AI-driven autonomy gains come at the cost of collective learning, or whether emerging practices such as AI-mediated collaboration introduce new forms of creative work.

Understanding how developers perceive and integrate AI into team-based work, especially in terms of creativity and collaboration, is an urgent research need. Such insight will clarify the limits and benefits of human–AI collaboration and guide design principles, ethical guidelines, and organizational practices that support effective, healthy teams and individuals in the age of AI. To address this gap, we conducted semi-structured interviews with software professionals to explore two research questions:

- \textbf{RQ1} How do software developers perceive and use AI tools in creative activities within software development teams?

- \textbf{RQ2} How does the use of AI tools influence collaboration and team dynamics in software development teams?

We report preliminary findings from interviews with 13 software professionals across four companies and various technical roles. AI supports creativity by helping developers explore the solution space, fill knowledge gaps, and surface previously unconsidered alternatives. Teams are testing new AI-mediated collaboration practices (e.g., using AI in group discussions and shared prompt writing) having reduced peer interaction and informal knowledge exchange.
\section{Background and Related Work}

Scholars consider creativity as related to novelty and ideas, such as in Boden's definition~\cite{boden2004creative}: \say{Creativity is the ability to come up with ideas or artifacts that are new, surprising, and valuable}. Software developers have a narrower focus and consider creativity as a form of problem-solving that emphasizes reuse and usefulness over originality and novelty~\cite{inman_developer_2024}. Popular creativity techniques used by developers include collaborative activities such as brainstorming and whiteboarding~\cite{groeneveld_creativity_2021}. Collaborative work is key for developers' creativity~\cite{inman_developer_2024} as is the ability to work independently, free from distractions, which facilitates the attainment of the flow state conducive to creativity~\cite{Ritonummi_flow_2023}. Along with a supportive work environment, technical skills, motivation for the task, and creativity-relevant processes also help individuals be creative at work~\cite{amabile_componential_model_2016}.

With the widespread usage of GenAI, several studies have examined its practical influence on developers' creativity, albeit indirectly. Two studies~\cite{jetbrains_Survey_2025, meem_ChatGPT_2025} exploring developers' use of GenAI have reported varying perceptions of its influence on creativity: some developers report that it has improved their creativity, whereas others report that it has worsened it. Relatedly, practitioners report that GenAI is rarely used for tasks traditionally considered creative, such as design~\cite{pereira_exploring_2025}. A recent interview study with 30 software development practitioners~\cite{zacharias2026developers} found that GenAI can stimulate creative problem-solving and enhance productivity, while also raising concerns about overreliance, skill degradation, and reduced communication between developers. In theory, GenAI has the potential for aiding creativity by quickly addressing knowledge gaps~\cite{pereira_exploring_2025}, supporting learning~\cite{meem_ChatGPT_2025}, and helping developers explore potential options~\cite{barke2023grounded}. Conversely, GenAI may inhibit creativity as developers ask GenAI for help rather than colleagues~\cite{salomon_GenAIHumanInteractions_2026}, reducing social interactions and their inherent learning opportunities. Also, senior developers have voiced concerns that novice developers are overly reliant on GenAI~\cite{pereira_exploring_2025}, which may hinder the development of core competencies critical to software developers, including the collaboration skills~\cite{zhou_developer_competencies_2018} that support creativity. Overall, evidence suggests that GenAI's effect on creativity depends on how critically developers engage with its outputs.

The effects of GenAI on team collaboration have also begun to attract research attention. Salomon et al.~\cite{salomon_GenAIHumanInteractions_2026} found that developers increasingly consult AI before asking colleagues, reducing interruptions but also weakening the peer exchanges through which tacit knowledge circulates, with colleagues being reserved for contextual reasoning and complex decisions. Seeber et al.~\cite{seeber2020machines} argue that for AI to act as a true team member, teams must rethink roles, orchestration, and shared situational awareness, which current tools only partly support. The combined impact of GenAI on creativity and collaboration in software teams remains underexplored. Prior studies tend to treat the two dimensions separately, leaving open how trade-offs between individual AI use and collective creative work manifest in practice. Our study differs from this prior work along three fronts. Salomon et al.~\cite{salomon_GenAIHumanInteractions_2026} examine changes in developer-to-developer interaction patterns but do not address creativity. Seeber et al.~\cite{seeber2020machines} offer a conceptual research agenda on AI as a teammate rather than empirical evidence from practitioners. Zacharias et al.~\cite{zacharias2026developers} report broad practitioner concerns about productivity, deskilling, and communication, but do not analyze creativity and collaboration jointly at the team level or examine AI’s role in collective creative work. 

\section{Research Method}

We conducted a qualitative interview study~\cite{seaman2008qualitative} to investigate how SE professionals perceive and integrate AI-assisted development tools into their daily practices, focusing on creativity and collaboration within teams. A qualitative design was selected to capture nuanced experiences, sensemaking processes, and contextual factors that quantitative measures alone cannot adequately address~\cite{seaman2008qualitative}.

We recruited 13 SE professionals (10 men and 3 women) from 4 companies (\autoref{table:participants}). Participants from the same company worked within the same team, providing both individual and within-team perspectives on AI use. Participants were required to: (i) have professional experience in software development and (ii) report active or recent use of AI-assisted tools (e.g., code assistants or generative AI systems) in collaborative development contexts. They were professionals from different roles (e.g., developers, designers, project managers), seniority levels, and organizational settings. For readability, we use the term \textit{developers} broadly throughout this paper to refer to the full spectrum of software professionals in our sample, including QA, project management, and leadership roles, except where a specific role is relevant to the discussion. We employed an iterative, non-probabilistic sampling strategy~\cite{baltes2022sampling}.We initially recruited participants through convenience sampling within our professional networks. As data collection and preliminary analysis progressed, we noted the absence of junior developers who began learning programming after the release of ChatGPT and then used purposive sampling to recruit this relevant subgroup\cite{baltes2022sampling}.

The study was approved by UFPE's ethics board. Participants were all volunteers and provided informed consent. All data were anonymized, treated confidentially, and stored securely.

\begin{table}[]\footnotesize
\caption{Interview participants profile}
\label{table:participants}
\begin{tabular}{llrc}
\textbf{ID} & \textbf{Current Role} & \textbf{Total experience (years)} & \textbf{Company} \\
\hline
P1  & Test team lead            & 14          & C1 \\
P2  & Software Engineer         & 7           & C1 \\
P3  & Software Engineer         & 8           & C1 \\
P4  & Project Manager           & 23          & C2 \\
P5  & Tech Lead                 & 13          & C1 \\
P6  & Tech Lead                 & 14          & C1 \\
P7  & Software Engineer Intern  & \textless 1 & C3 \\
P8  & Software Engineer Intern  & 1           & C3 \\
P9  & Software Engineer         & 2           & C2 \\
P10 & QA Engineer               & 10          & C4 \\
P11 & QA Lead                   & 26          & C4 \\
P12 & Software Engineer         & 5           & C2 \\
P13 & Software Engineer         & 5           & C2
\end{tabular}
\end{table}

\subsection{Data Collection and Analysis}

Data were collected through semi-structured interviews conducted online via Google Meet in February 2026. Each session ranged from 45 to 60 minutes and was audio-recorded with participants' consent. The interview protocol~\footnote{\url{https://doi.org/10.6084/m9.figshare.31489231}} included open-ended questions addressing: integration of AI tools into development workflows; perceived impact on creativity and ideation; effects on team coordination and shared understanding; and perceived risks, limitations, and organizational implications. The protocol was piloted with 2 participants; based on their feedback, we adjusted question wording and these pilot interviews were discarded from the final dataset. The protocol served as a flexible guide rather than a rigid script, but interviewers ensured that all planned topics were covered to avoid omissions. Interviews were transcribed verbatim, with participants having pseudonyms assigned and identifying information removed.

We analyzed the data using inductive thematic analysis~\cite{braun2006using}. Two researchers independently open-coded the interview transcripts and regularly compared interpretations, resolved disagreements, and refined the coding structure. Codes were then grouped into higher-level themes through a researcher-led process supported by an LLM, used only to suggest alternative groupings and exploratory interpretations. All analytical decisions remained the responsibility of the research team. The use of LLMs to support qualitative analysis in SE research has been discussed recently, provided that human judgment governs analytical decisions~\cite{trinkenreich2025get, bano2024large}.As coding progressed and themes emerged, peer debriefing with a third researcher was used to critically examine interpretations, challenge assumptions, and refine themes, strengthening the study’s credibility and trustworthiness. Coding and theme development used shared documents to ensure transparent, traceable analytic decisions.

\section{Findings}

\begin{figure}
    \centering
    \includegraphics[width=1\linewidth]{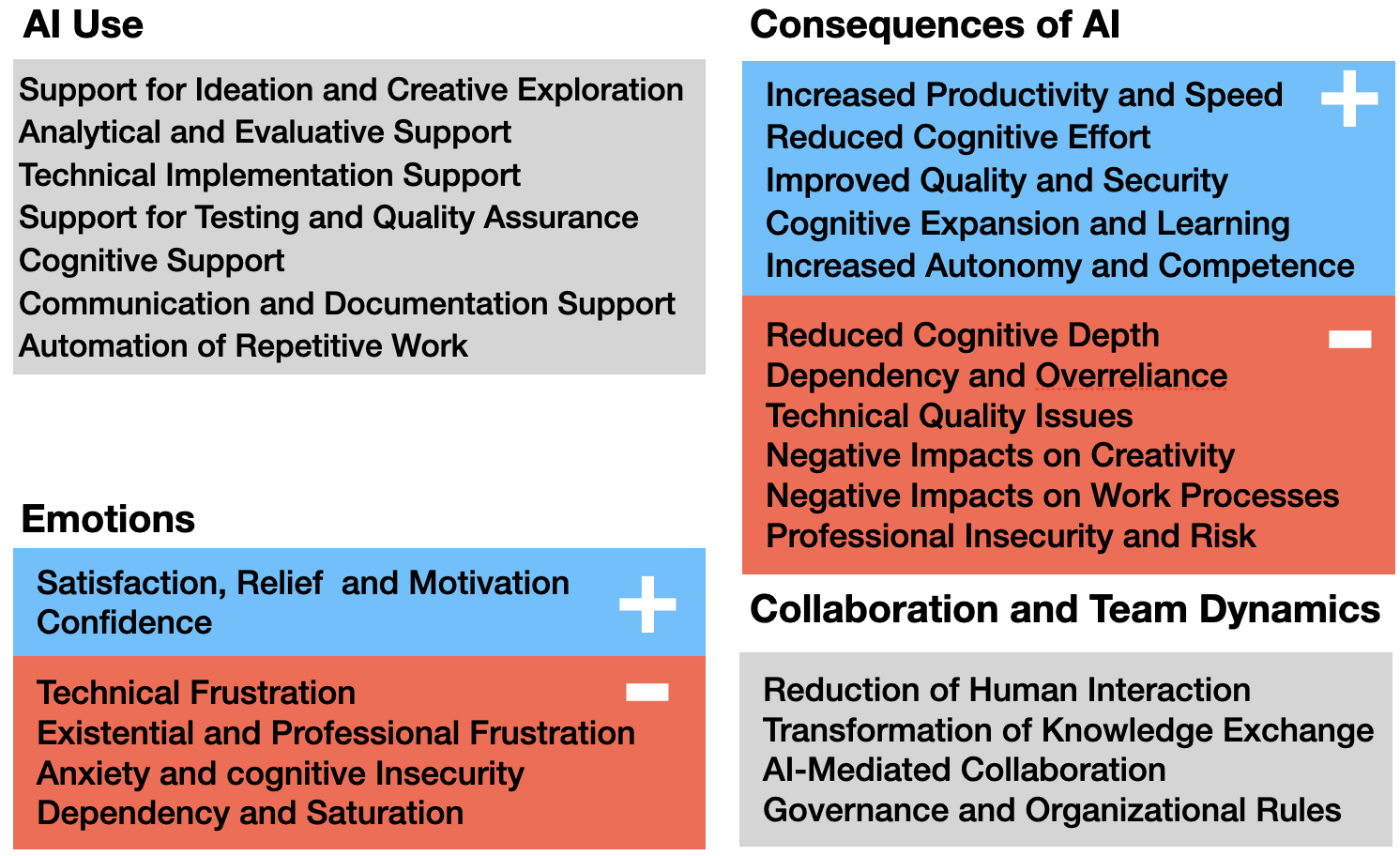}
    \caption{Themes and categories of the thematic analysis. Collaboration and Team Dynamics are presented without positive/negative distinctions, unlike AI Use and Consequences of AI, where participants reported directional effects.}
    \Description[Shows the four themes and categories within the themes]{Shows the four themes of AI Use, Emotions, Consequences of AI, and Collaboration and Team Dynamics. Within these, numerous categories are shown that are subsequently described in Section 4.}
    \label{fig:categories}
\end{figure}

The two RQs addressed, respectively, the impacts of AI on creativity and on collaborative work in software teams. \autoref{fig:categories} summarizes the themes and categories identified. Our analysis revealed three central themes: (1) AI Use; (2) Consequences of AI; and (3) Collaboration and Team Dynamics. \textit{Emotions} emerged as a cross-cutting dimension: emotional responses were not confined to a separate domain but arose in direct connection with creative and collaborative experiences, shaping how participants made sense of both. 

\subsection{Thematic Analysis Results}

\subsubsection{AI Use}
\label{sec:aiusage}
This theme captures how developers integrate AI tools into their workflows. Participants described AI not only as a coding assistant but as a cognitive, creative, analytical, and communicative resource. AI was used for ideation, technical implementation, documentation and automation, indicating that AI tools are a multi-functional component embedded in daily development practices. These patterns suggest that AI is being appropriated across the full breadth of development work, including activities with creative components, rather than routine or purely technical tasks.

\noindent\textbf{Support for Ideation and Creative Exploration.}
Participants described using AI to stimulate alternative solutions, support architectural brainstorming, and generate proofs of concept. AI was frequently used to explore design possibilities and iteratively refine ideas. While some perceived AI as expanding creative exploration, others noted that ideas sometimes ``arrive already formed,'' potentially constraining deeper ideation. For example, P6 described using AI to generate multiple solutions and evaluate the cost-benefit of each before selecting one: \say{\textit{Trying to explore possible solutions, what ends up happening is a list of options and the cost-benefit of each one...to explore options that we haven't even considered yet.}}

\noindent\textbf{Analytical and Evaluative Support.}
Participants emphasized the importance of critically evaluating AI outputs. Responses were compared, cross-validated, and reviewed by experienced developers. AI was not perceived as a source of truth, but as a hypothesis generator requiring domain knowledge and human judgment: \say{\textit{Sometimes it suggests things, we review the document and think, `Oh no, what it said doesn't make any sense. Let's remove it.'}} (P4)

\noindent\textbf{Technical Implementation Support.}
AI was used in everyday development tasks, particularly for generating code snippets and simplifying implementations. Participants described AI as an assistant, intern, or partner capable of accelerating implementation while developers retained responsibility for final decisions. As noted by P13: \say{\textit{The coding, well, I think 70\% is AI, 30\% is me. I just give some pointers, see something that can be done more efficiently.}}

\noindent\textbf{Support for Testing and Quality Assurance.}
Participants noted that AI makes it easier to create tests, including integration tests (P1), but expressed caution in security-sensitive or quality-critical contexts. AI-generated unit tests and review suggestions were considered limited without sufficient contextual knowledge, with senior developers viewed as essential for final validation.

\noindent\textbf{Cognitive Support.}
AI functioned as an on-demand tutor, helping developers overcome knowledge gaps and clarify doubts, \say{\textit{shortening the learning curve}} (P12) when learning new topics. However, effective use required knowing how to write prompts and interpret responses critically. P7 noted they asked AI to explain unfamiliar generated code: \say{\textit{I either ask it to explain it better or I also use the chat to explain what's going on.}}

\noindent\textbf{Communication and Documentation Support.}
AI was used to improve language clarity, draft documentation, summarize materials, assist in English communication, and support discussions. Notably, some participants reported that interacting with AI had become faster than consulting colleagues for these tasks, a pattern with direct implications for peer interaction within teams. Documentation tasks ranged from summarizing requirements to authoring commit messages generated entirely by AI (P13).

\noindent\textbf{Automation of Repetitive Work.}
Participants reported using AI to optimize workflows and automate repetitive tasks (P4). At the same time, they acknowledged the risks of excessive reliance, particularly for trivial activities that may erode manual proficiency. P3 worried that such a \say{\textit{strong bond}} had formed that removing AI would drastically reduce their productivity.

\subsubsection{Consequences of AI}
\label{sec:consequences}
Beyond usage patterns, participants reflected on the broader cognitive, professional, and creative consequences of AI integration. AI tools were associated with productivity gains and increased autonomy, but also raised concerns about reduced analytical depth, dependency, and potential impacts on creativity and expertise development, illustrating a duality of AI as both an enabler and a potential disruptor of professional practice. Positive (+) and negative (-) subcategories are indicated below.

\noindent\textbf{(+) Increased Productivity and Speed.}
AI was seen to speed up tasks and boost efficiency, especially for implementation-heavy work, as P5 noted: \say{\textit{[AI] autocomplete helps a lot.}}

\noindent\textbf{(+) Reduced Cognitive Effort.}
AI reduced the need to interrupt colleagues and provided rapid answers to technical questions. This improved workflow continuity, but some participants expressed concern about reduced cognitive engagement. P7 reframed this as a positive contribution to team dynamics: \say{\textit{I see it a lot like, in a way, I'm freeing up the time and energy of these people who spend more time there, things that are often basic and just that.}}

\noindent\textbf{(+) Improved Quality and Security.}
Some participants perceived quality gains when AI suggestions were carefully validated, though improvements were conditional on human oversight and domain expertise. P1 noted that AI \say{\textit{provides a bit better coverage than the developer might have considered before}}, while still acknowledging ongoing concerns about security issues such as SQL injection.

\noindent\textbf{(+) Cognitive Expansion and Learning.}
AI facilitated exposure to alternative approaches and supported learning through iterative exploration. P5 described using AI to acquire targeted knowledge before committing to a solution: \say{\textit{I can ask specific questions so I can understand what it's for, in what contexts that technology or approach can or cannot solve a problem. And then I decide what is worth dedicating my effort to.}}

\noindent\textbf{(+) Increased Autonomy and Competence.}
Participants reported greater independence, requiring less assistance from teammates. AI enabled more autonomous problem-solving: \say{\textit{we can have more autonomy to solve things, and it eliminates several bottlenecks, even in terms of learning.}} (P12)

\noindent\textbf{(-) Reduced Cognitive Depth.}
Several participants raised concerns about diminished analytical engagement and superficial understanding of solutions, particularly among less experienced developers. P9, a novice developer, felt that increasingly capable AI tools could lead developers to become \say{\textit{a little lazy sometimes.}}

\noindent\textbf{(-) Dependency and Overreliance.}
Overdependence on AI was reported as a risk, including difficulty reverting to manual practices and incomplete understanding of generated solutions. P6 described the phenomenon directly: \say{\textit{Today it's almost become an addiction, you know, to do everything with artificial intelligence.}}

\noindent\textbf{(-) Technical Quality Issues.}
AI-generated errors, hallucinations, and lack of contextual awareness were observed. Undetected issues could delay delivery or compromise architectural decisions. P5 cautioned: \say{\textit{It might provide information that isn't 100\% accurate.}}

\noindent\textbf{(-) Negative Impacts on Creativity.}
While AI supported idea generation, some participants believed that excessive reliance weakens originality and deeper creative reasoning. P6 observed: \say{\textit{It doesn't let things flow like they did before using artificial intelligence.}}

\noindent\textbf{(-) Negative Impacts on Work Processes.}
AI changed workflow dynamics, shifting how teams coordinate tasks and manage responsibilities. P1 noted reduced involvement in discussions: \say{\textit{I don't really see that interaction between people. It's just one or two people who are there.}}

\noindent\textbf{(-) Professional Insecurity and Risk.}
Participants had concerns about generational skill differences, erosion of expertise, and potential long-term career impacts. P7 warned: \say{\textit{If you don't think, don't ask questions, don't read what AI is doing, I think AI can harm a career.}}

\subsubsection{Emotional Responses to AI}
AI tool usage generated varied emotional responses. Participants described feelings ranging from satisfaction (e.g., P1, P5) and relief (e.g., P3) to frustration (e.g., P1, P5), anxiety (e.g., P13), and professional insecurity (e.g., P3, P11). These affective reactions were closely linked with perceptions of control, competence, trust, and dependency, highlighting the emotional dimension of AI-mediated work.

\noindent\textbf{(+) Satisfaction, Relief, Motivation, and Confidence.}
AI's speed in completing tasks generated relief and satisfaction, even in tasks where developers lacked expertise (P10). For some, AI served as a supportive partner that increased confidence to assume more responsibilities: \say{\textit{We end up trusting ourselves to take on more responsibilities because, before, we had all this uncertainty.}} (P6). P2 said that AI-enabled overdelivery against managerial expectations boosted motivation: \say{\textit{Motivation comes from a consequence.}}

\noindent\textbf{(-) Technical Frustration.}
Similar to other studies~\cite{pereira_exploring_2025}, hallucinations, verbosity, and unreliable outputs caused frustration, especially when correction effort was required. P2 described the experience: \say{\textit{It is the same as talking to a wall when the AI is hallucinating.}}

\noindent\textbf{(-) Existential and Professional Frustration.}
Some participants expressed concerns about the erosion of professional identity. P6 felt \say{\textit{not needed}}, with the leading role falling to AI. P12 felt their skills were no longer required, as AI goes straight to the solution, leaving them feeling less \say{\textit{active in achieving the desired result}}. These concerns extended to fears of being replaced (P13).

\noindent\textbf{(-) Anxiety and Cognitive Insecurity.}
Insecurity emerged when AI-generated solutions were not fully understood. P5 worried that AI would \say{\textit{probably lead to solutions that you won't even understand}}, while P13 feared forgetting how to work without AI support.

\noindent\textbf{(-) Dependency and Overuse.}
Some participants described fatigue associated with intensive AI interaction. P6 reported feeling isolated from colleagues due to AI-only workflows. P11 expressed concern about no longer trusting their own unassisted work: \say{\textit{no longer trust what I do on my own.}}

\subsubsection{Collaboration and Team Dynamics}
\label{sec:collaboration}
AI integration extended beyond individual cognition, reshaping team-level dynamics. Participants reported shifts in communication patterns, knowledge exchange, and collaboration structures, including increased individualization of work and reduced peer interaction. At the same time, teams experimented with AI-mediated collaboration and established governance mechanisms to regulate its use.

\noindent\textbf{Reduction of Human Interaction.}
AI consultation sometimes replaced peer collaboration, leading to more individualistic workflows. P8 described the shift: \say{\textit{Instead of asking my colleagues for advice, I end up asking the AI and researching, creating a flow there that's mostly on my own.}}

\noindent\textbf{Transformation of Knowledge Exchange.}
AI altered how experience and tacit knowledge circulate within teams. Prompt engineering has become a shared practice, but some participants worried that mentoring and contextual knowledge transfer have weakened. P1 reflected: \say{\textit{The exchange of experiences I had in the past with code [...] I see that this has diminished; this learning, this searching, this exchange between people makes us learn much more. And I don't see that much anymore today.}}

\noindent\textbf{AI-Mediated Collaboration.}
Teams experimented with collaborative prompt writing, AI-assisted discussions, and AI-integrated pair programming. P2 described how their team incorporates AI into group conversations: \say{\textit{Usually, it's people chatting and someone says, ``No, wait a minute, guys, I'm going to create a GPT chat here, let's see what it says." That's the reality of how it works.}} This represents an emerging triadic collaboration pattern involving developers, colleagues, and AI, in which the tool participates in collective reasoning rather than serving individual workflows alone.

\noindent\textbf{Governance and Organizational Rules.}
Organizations implemented varying levels of formal governance, including usage policies, experimentation strategies, and tool standardization. P9 described a structured approach: \say{\textit{We have some pre-established commands with pre-made, reviewed templates [...] before I send it for review by other humans, I already have it reviewed by AI.}}

\subsection{Answers to Research Questions}

\noindent\textbf{RQ1: How do software developers perceive and use AI tools in creative activities within software development teams?}
We interpret creative activities broadly, encompassing not only ideation and design but also the exploratory and problem-solving dimensions of everyday development work (e.g., testing, documentation, automation), since participants described these tasks as involving choices among alternative approaches rather than purely mechanical execution. Participants reported using AI across multiple stages of the development process, including activities with creative components. AI was frequently used to support ideation and exploration of alternative solutions, particularly during architectural discussions and prototyping (see Section~\ref{sec:aiusage}). It was also used for implementation, analytical tasks such as comparing approaches, cognitive tasks such as learning unfamiliar technologies, and communication tasks such as drafting documentation. Developers described AI as a flexible tool supporting both exploration and execution. However, they consistently emphasized the need for human validation and critical evaluation, while warning that heavy reliance may reduce independent ideation over time (Section~\ref{sec:consequences}).

\noindent\textbf{RQ2: How does the use of AI tools influence collaboration and team dynamics in software development teams?}
Participants reported that AI reshapes how developers interact and collaborate within teams. A recurring observation was that developers increasingly consult AI instead of asking colleagues, reducing interruptions but also decreasing opportunities for peer discussion and informal learning (see \textit{Reduction of Human Interaction} and \textit{Transformation of Knowledge Exchange}, Section~\ref{sec:collaboration}). At the same time, some teams developed emerging triadic collaboration patterns involving developers, colleagues, and AI, such as shared prompt writing and integrating AI into group discussions (see \textit{AI-Mediated Collaboration}, Section~\ref{sec:collaboration}). AI also enabled greater individual autonomy, allowing developers to solve problems without relying on teammates (see \textit{Increased Autonomy and Competence}, Section~\ref{sec:consequences}). Despite these shifts, participants consistently positioned humans as central to development teams: architectural decisions, code validation, and leadership responsibilities were described as human tasks, with AI perceived as a supporting resource rather than a replacement for human expertise (see Section~\ref{sec:aiusage}).

\section{Discussion}

Our findings indicate that AI is operating simultaneously as an amplifier of individual capability and as a fragmenting force on collective work practices. Developers report gains in autonomy, speed, and creative exploration, but these benefits come with observable costs to peer interaction, mentoring, and shared knowledge. We discuss the implications of these tensions below.

\noindent\textbf{Emerging Triadic Collaboration Patterns.}
A finding with limited precedent in prior literature is the emergence of triadic collaboration patterns involving developers, colleagues, and AI acting jointly in team settings. Participants described practices such as collectively prompting AI during group discussions and using AI outputs as shared reference points for decision-making. These practices differ qualitatively from individual AI use and from traditional human-human collaboration, suggesting a new mode of collective work that existing frameworks for team collaboration~\cite{seeber2020machines} do not yet fully account for. Understanding how these patterns develop, stabilize, and affect team cohesion and knowledge sharing is an important direction for future research.

\noindent\textbf{Asymmetric Human--AI Collaboration.}
Participants described AI as generating ideas, code, and explanations, while humans remained responsible for evaluating outputs. Developers framed AI as proposing alternatives rather than making decisions, resulting in a collaboration where AI provides generative support, and humans retain oversight and accountability. This aligns with human-centered AI principles~\cite{amershi2019guidelines} and with findings by Salomon et al.~\cite{salomon_GenAIHumanInteractions_2026}, about developers using AI for routine technical questions while relying on colleagues for contextual reasoning and complex decisions. Our data extend this picture by showing that asymmetry also manifests in creative activities: AI generates options, but the evaluative and selective dimensions of creative work remain human.

\noindent\textbf{AI Primarily Supporting the Solution Space.}
Participants reported using AI mainly for implementation tasks such as generating code, exploring algorithms, summarizing documentation, and debugging. In terms of the Double Diamond model~\cite{doublediamond}, these uses fall primarily within the \emph{solution space}. AI was rarely mentioned in activities related to the \emph{problem space}, such as identifying user needs or framing requirements, possibly reflecting the participants' roles as developers rather than product managers or designers~\cite{gama_Startups_Creativity_2025}. This asymmetry suggests that AI may reinforce existing divisions of creative labor in software teams rather than redistribute them.

\noindent\textbf{Changes in Perceptions of Creativity.}
Our results indicate that AI may be shifting how developers understand creativity in software development. Participants described creativity not only as generating new ideas but also as selecting, refining, and improving alternatives. Because AI can generate multiple candidate solutions quickly, developers may increasingly engage in evaluative forms of creativity, where the creative task involves selecting and adapting AI-generated options rather than producing ideas from scratch. This shift resonates with Amabile's componential model~\cite{amabile_componential_model_2016}, which distinguishes generative from evaluative creative processes: AI appears to be compressing the generative phase while expanding the evaluative one. Yet, some participants suggested that heavy reliance on AI could reduce opportunities for independent ideation~\cite{gama_Startups_Creativity_2025}, a tension that warrants further empirical investigation.

\noindent\textbf{Potential Deskilling Effects.}
Some participants expressed concern that frequent reliance on AI-generated solutions may reduce the need for deep reasoning or encourage superficial understanding of generated code, particularly among less experienced developers. Similar risks have been discussed in automation research~\cite{parasuraman1997humans}, where extensive reliance on automated systems reduces opportunities to develop expertise. Zacharias et al.~\cite{zacharias2026developers} report convergent findings from a broader practitioner study, documenting concerns about skill degradation and an ``illusion of competence'' among novice developers who mistake AI-generated outputs for genuine understanding. However, other participants in our study described AI as a learning aid that helps them understand unfamiliar technologies or explore alternative approaches. This contrast suggests that deskilling is not inevitable but may depend on whether developers engage critically with AI outputs or defer to them uncritically.

\noindent\textbf{Erosion of Social Learning Dynamics.}
Participants reported that AI increasingly replaces situations in which developers previously consulted colleagues. Salomon et al.~\cite{salomon_GenAIHumanInteractions_2026} reported a similar pattern, where developers consult AI before asking colleagues, reducing interruptions but potentially altering knowledge exchange within teams. From the perspective of Bandura's social learning theory~\cite{bandura1977social}, learning often occurs through observation and interaction with others. If AI substitutes for some of these interactions, opportunities for informal knowledge exchange may decrease. Our data add a further dimension: participants reported not only reduced peer consultation but also weakened mentoring relationships, suggesting that the effects on collective learning may be more structural than previously documented. Chen and Lee~\cite{chen2026effects} demonstrate that team self-efficacy and team identification are prerequisites for failure-based learning behavior, which fully mediates team performance in adaptive development contexts. This is the kind of interaction AI is displacing, risking the ``illusion of competence''~\cite{zacharias2026developers} at scale.

\noindent\textbf{Effects on Collective Creativity.} Beyond individual learning, this reduction in peer interaction may also suppress collective creativity. Hargadon and Bechky~\cite{hargadon2006collections} argue that creative solutions often emerge not from individual effort but from momentary collective processes, in which social interactions activate knowledge associations that no individual could reach alone. Practices such as asking a colleague for help, explaining a problem out loud, or observing how others approach a task are precisely the interactions their framework identifies as generative. This perspective aligns with Amabile's argument that failures are learning experiences that can lead to creative and successful outcomes~\cite{amabile_componential_model_2016}. In collaborative work, when teams face failed ideas and misunderstandings, this creates opportunities for shared reflections and creative recombination. If AI is now absorbing these interactions, it may be displacing not only knowledge exchange but also the social contexts from which collective creativity arises. This raises an important open question for future research: whether AI is genuinely substituting for collective creativity, or merely shifting the form it takes, as suggested by the triadic collaboration patterns observed in this study.

\noindent\textbf{Preservation of Human Agency.}
Despite growing use of AI, participants consistently positioned humans as responsible for decision-making, code validation, and architectural reasoning. This active preservation of human agency can be understood as a response to the deskilling concerns and collective learning losses described above: by maintaining ownership of high-stakes decisions, developers appear to be negotiating a boundary that protects both individual expertise and team accountability. This informal norm (positioning AI as a supporting tool rather than an autonomous decision-maker) aligns with recent findings~\cite{salomon_GenAIHumanInteractions_2026} and with human-centered AI design principles~\cite{amershi2019guidelines}, and may be worth formalizing in organizational governance frameworks.

\section{Limitations and Threats to Validity}

This study is subject to social desirability bias and selective recall when participants report their GenAI usage. Recurring themes across participants reduce, though do not eliminate, these risks. Researcher subjectivity in analysis and interpretation is also a concern; to mitigate this, the analytical process was documented and the research team comprises SE academics with industry experience. This was a single country study, and findings may not transfer to contexts with different cultural or organizational norms around AI adoption. The sample included only technical staff, excluding roles such as product managers who may engage with AI differently in creative work. Also, although participants from the same company worked within the same team, our data collection was based on individual interviews instead of team-level observation; some findings framed at the team level (e.g., collaboration and governance patterns) are therefore inferred from aggregated individual perceptions rather than directly observed shared team practices. As a preliminary study, no claims to thematic saturation are made.

\section{Conclusion and Future Work}

This preliminary study investigated how AI tools influence creativity and collaboration in software development teams. Developers integrate AI across multiple stages of the development process, using it as a resource for ideation, exploration of alternatives, implementation, and communication. At the same time, participants reported tensions: reduced peer interaction, risk of overreliance, concerns about the erosion of critical thinking, and weakened mentoring relationships. Taken together, the findings suggest that AI reshapes both creative practices and collaborative dynamics rather than simply improving individual productivity.

Four directions emerge for future work. First, longitudinal studies should examine whether AI-related deskilling accumulates or diminishes over time. Second, the triadic collaboration observed among developers, colleagues, and AI warrants investigation into how teams establish shared norms, roles, and situational awareness. Third, future research should explore why AI use remains concentrated in solution development rather than problem framing and requirements engineering, and how organizational factors shape this distribution. Fourth, the decline in peer interaction raises the question of whether AI suppresses collective creativity~\cite{hargadon2006collections} or enables new forms of AI-mediated collaboration. Longitudinal and observational studies could distinguish between these possibilities.


\section*{Artifact Availability}

The interview protocol is publicly available at \url{https://doi.org/10.6084/m9.figshare.31489231}. Interview transcripts are not shared to protect participant confidentiality, aligned with the study's ethics approval.

\bibliographystyle{ACM-Reference-Format}
\bibliography{references}

\end{document}